# Optimized Feature Selection and Neural Network-Based Classification of Motor Imagery Using EEG Signals

Muhammad Sudipto Siam Dip[1], Mohammod Abdul Motin[1*], Md. Anik Hasan[1], and Sumaiya Kabir[1,2],

[1]Department of Electrical & Electronic Engineering,
Rajshahi University of Engineering & Technology, Rajshahi-6204, Bangladesh.
[2]School of Engineering, RMIT University, Victoria, Australia.
*Corresponding Author: m.a.motin@ieee.org

**Abstract—Objective:** Machine learning- and deep learning-based models have recently been employed in motor imagery intention classification from electroencephalogram (EEG) signals. Nevertheless, there is a limited understanding of feature selection to assist in identifying the most significant features in different spatial locations. **Methods:** This study proposes a feature selection technique using sequential forward feature selection with support vector machines and feeding the selected features to deep neural networks to classify motor imagery intention using multi-channel EEG. **Results:** The proposed model was evaluated with a publicly available dataset and achieved an average accuracy of 79.70 ± 7.98% for classifying two motor imagery scenarios. **Conclusions:** These results demonstrate that our method effectively identifies the most informative and discriminative characteristics of neural activity at different spatial locations, offering potential for future prosthetics and brain-computer interface applications. **Significance:** This approach enhances model performance while identifying key spatial EEG features, advancing brain-computer interfaces and prosthetic systems.

## I. Introduction

Motor imagery is a process that captures brain activity and analyses it to interpret user actions. It is a spontaneous potential generated by the brain without external stimulus, with potential applications in brain-computer interfaces (BCIs), medical rehabilitation, brain-controlled devices, and prosthetic hands [1-3]. Since the electroencephalogram (EEG) is non-invasive and used in several types of brain applications, it is continuously gaining attention in developing different applications of BCIs, such as motor imagery classification [4]. However, the non-stationary nature of the brain signal picked up by the EEG makes it hard to recognize patterns in the signal and classify them as motor imagery [5].

The automated methods for EEG-based motor imagery classification are broadly divided into two categories: those using handcrafted features combined with traditional machine learning classifiers such as support vector machines (SVM), linear discriminant analysis (LDA), k-nearest neighbors, and those employing neural network-based automated feature extraction and classifiers [6]. The conventional features of EEG for motor imagery classification includes spatial features [7-9], frequency-spatial features [10-12], and temporal-frequency features [13, 14]. Among spatial feature extraction techniques common spatial patterns (CSP) and independent component analysis aim to enhance motor imagery-related features by extracting spatial patterns and separating EEG sources [15]. Moreover, different statistical and entropy based features were previously reported [16], [17], [18].

Apart from the hand-crafted features with traditional classifiers, different researchers have also explored advanced deep learning-based frameworks [21-25]. However, direct input of raw EEG signals often suffers from several issues, including low signal-to-noise ratio, inherent non-stationarities, and baseline problems [19], and have limited interpretability. To overcome these problems, different signal processing techniques were applied to reduce or partially remove those noises before feature extraction and fed features to deep learning models as classifiers. Among different preprocessing techniques, bandpass filtering, independent component analysis, artifact subspace reconstruction, etc., were applied to improve signal quality [20, 21].

In [22], R. Corralejo *et al.* incorporated a genetic algorithm (GA) with different types of features such as continuous and discrete wavelet transforms, spectral features, and also other statistical techniques such as an autoregressive model and μ based rhythm matched filter. However, due to GA's stochastic (random) nature, it can get stuck in what may be a good set but not necessarily the best/optimal set of features [23]. Xinyang Yu *et al.* used principal component analysis (PCA) as their feature reduction technique with spatial filter-based feature extraction methods [24]. However, PCA focuses on variance to reduce dimensionality and doesn't necessarily select the most discriminative features when classifying different motor imagery tasks. The sequential forward floating selection (SFFS) method has previously been used in channel selection for motor imagery-based BCI [25]. M. H. Bhatti *et al.* used sequential backward floating selection, an alternative version of SFFS, which provides effective feature selection with RBFNN classifier and performs better than recursive feature elimination (RFE) with SVM classifier [26]. A. Liu *et al.* showed that the firefly algorithm can be utilized for motor imagery classification by selecting the best subset of features [27]. They implemented spectral regression discriminant analysis (SRDA) as their classifier and obtained an average accuracy of 70.20%. In [28], NS Malan and S Sharma introduced regularized neighborhood component analysis (RNCA) for motor imagery classification with an SVM classifier where RNCA with SVM outperformed ReliefF, GA, and PCA, but at the same time, RNCA required multiple parameters to be optimized to generalize the model. Jing Jiang *et al.* compared four different feature selection technique, namely mutual information, least absolute shrinkage and selection operator (LASSO), PCA, and steps-wise linear discriminant analysis, which were incorporated with CSP-based feature extraction. Among these, LASSO with the SVM achieved the best performance with 88.58% average accuracy [29]. Further, Yao Guo *et al.* used the minimal-redundancy-maximal-relevance (MRMR) technique to eliminate features that were either irrelevant or redundant from the overall feature subset. Their method used an LDA classifier to enhance the classification performance [30].

Most of these previous studies mainly focused on improving motor imagery classification using existing feature selection techniques. [ref] While, few have explored efficient methods that target optimal information for the learning algorithm. [ref] As feature number increases, dimensionality issues arise and computational complexity increases [ref]. The hypothesis is that employing a large number of relevant features specific to specific channels can resolve these issues without compromising accuracy. In this work, we extracted hand-crafted features from the statistical time domain, spectral domain, and time-frequency from EEG. We propose a hybrid feature selection technique based on modified SFFS to find out the optimal number of features from each channel to be fed into the neural network for EEG-based motor imagery classification. This paper is organized as follows: Section II presents our overall approach in a nutshell and details of our strategy. Section III describes the experimental result and evaluation. Section IV presents a brief analysis/outcome of the feature selection approach. Finally, the conclusion is given in section V.

## II. DATASET AND PROPOSED FRAMEWORK

### A. Dataset

In this work, we used the publicly available BCI competition 2008 – Graz dataset [31]. This dataset contains 9 participants in total. The experimental setup for this dataset contains 22 channels of EEG with the left mastoid as a reference. As our analysis aimed for EEG signals, we only used the EEG channel recordings. The dataset consists of recordings from two sessions for each subject, with six runs per session, a total of 12 runs for each participant. Each run contains 48 trials, with 12 trials for each class, resulting in 288 trials in a single session. The sampling frequency for the recorded signal was 250Hz. A bandpass filter was applied between 0.5Hz and 100Hz, with a 50Hz notch filter used to remove line noise. The amplifier voltage is constantly maintained at 100 microvolts during the session. The recording paradigm is illustrated in Fig. 1.

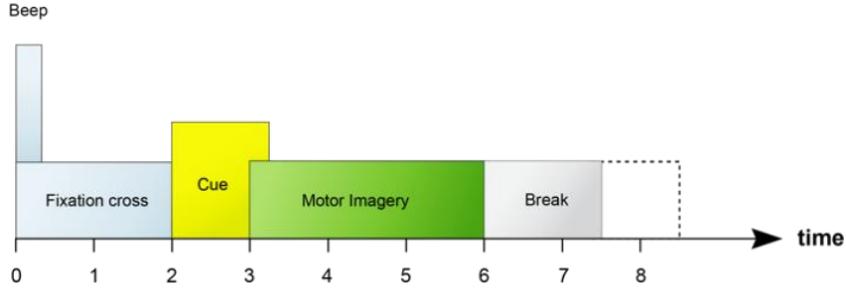

Fig. 1. Experimental paradigm: Participants were instructed to complete the motor imagery task until the fixation cross disappeared from the screen at 6 seconds.

### B. Proposed Model

The block diagram of the proposed model is illustrated in Fig. 2. The proposed model consists of four parts: i) Preprocessing, ii) Feature extraction, iii) Feature selection, and iv) Model architecture and hyper parameter tuning.

#### 1) Preprocessing

A finite impulse response bandpass filter was applied between 0.5Hz and 35Hz. The artifact subspace reconstruction method was employed to clean up the noisy portion of the data [20]. This process involved identifying and eliminating segments that were either discontinuous or noisy. Furthermore, the utilization of the independent component analysis technique was employed in order to identify and classify components that are associated with eye movements and muscle artifacts. EEGLAB [32] is used to preprocess the signal for these stages. Previous studies have demonstrated that the duration of motor imagery does not exceed 0.5 seconds [33]. We extracted epochs allied with the target label. These epochs were selected 0.2 seconds before the event onset and finished 0.5 seconds after the event onset for all the events, as shown in Fig. 3. To capture the pre-stimulus activity, the additional 0.2 seconds before the event initiation was included in our study. This uniform temporal segmentation allowed us to capture any subtle pre-event neural activity preceding the actual stimulus.

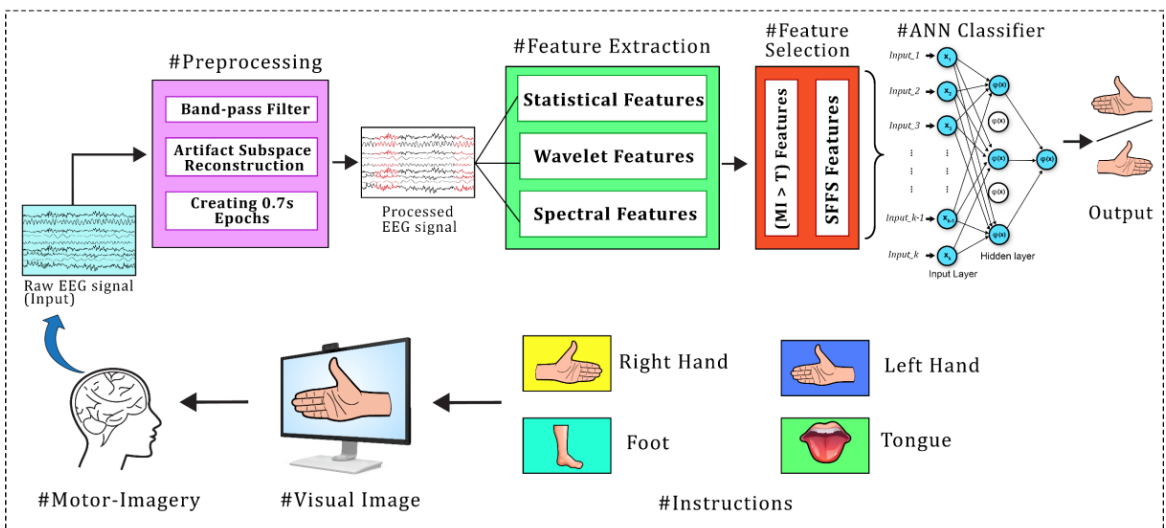

Fig. 2. The overall framework of our approach is divided into two sections. Firstly, with the raw dataset preprocessing takes place to improve the signal. Unwanted channel removal, filtering the signal to keep within low frequency band region and Artifact Subspace Reconstruction (ASR) to improve signal quality and finally uniform temporal segmentation to create 3D dataset. In the second section, the signal then separated for validation, one for training the model and another is for evaluation. Feature extraction and standardization is applied to both. The two-steps feature selection takes place and then with the optimal set a neural network is constructed, trained and evaluated on the validation set with the same optimal feature set selected.

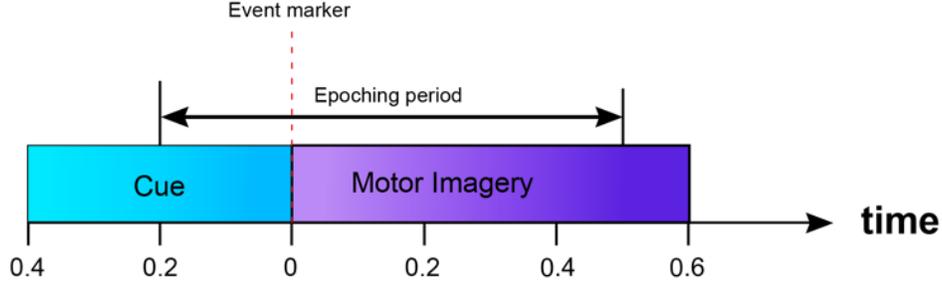

Fig. 3. Uniform temporal segments of EEG data before and after stimuli onset.

2) *Feature Extraction*

A set of 418 distinctive features from statistical time domain, spectral, and time-frequency domain were extracted from 22-channel EEG. The feature extraction pipeline encompassed three categories. In the first category, the power spectral density features using the Welch method [35] were explored. The spectral power distribution in delta, theta, alpha, low beta, mid beta, and beta was computed. The frequency range for each band is shown in Table I. The calculation of the power spectral density of an EEG epochs occurred in four steps which includes i) overlapping segmentation ii) applying window function iii) computing periodogram iv) average of periodograms.

Let's assume $x[n]$ is an EEG epoch of length $N$ into $K$ overlapping segments where each segment is of length $L$, and the overlap between segments is $D$. Then, a window function $w[m]$ is applied to each segment. Here, the $k^{th}$ segment of the signal is:

$$x_k[n] = x[n + kD].w[n] \quad (1)$$

For $n=0, 1 \ldots L-1$ and $k=0, 1, \ldots K-1$

Then, the periodogram for each of the segments was computed by doing the discrete Fourier transform of each windowed segment with this:

$$P_k(f) = \frac{1}{L \sum_{n=0}^{L-1} w[n]^2} \left| \sum_{n=0}^{L-1} x_k[n] e^{-\frac{j2\pi f n}{L}} \right|^2 \quad (2)$$

Where, $f$ is considered as the frequency bin. The average periodogram of the $K$ segments was computed to get the Welch's power spectral density estimate:

$$\hat{P}(k) = \frac{1}{K} \sum_{k=0}^{K-1} P_k(f) \quad (3)$$

In the second category of feature sets, the time-frequency domain features were extracted using wavelet transform. The Morlet mother wavelet was used to extract wavelet coefficient features. The mother wavelet was utilized by the function:

$$\Psi(t) = \pi^{-\frac{1}{4}} e^{i\omega_0 t} e^{-\frac{t^2}{2}} \quad (4)$$

The Morlet mother wavelet was used to decompose each channel's signal into a set of wavelet coefficients that represent the signal using the equation below:

$$W(a, b) = \frac{1}{\sqrt{a}} \int_{-\infty}^{+\infty} x(t) \psi^* \left( \frac{t-b}{a} \right) dt \quad (5)$$

Where, $W(a, b)$ is the wavelet coefficient at scale $a$ and translation $b$, $x(t)$ is the input signal, and $\Psi(t)$ is the Morlet wavelet function.

To enhance the relevancy of these coefficients, we compressed them using PCA into six distinct components that encapsulated the essence of the original wavelet coefficients. An array of 6 scales for the CWT, ranging from 1 to 128 on a logarithmic scale with a base of 2 was computed. The features from the first principal component were calculated.

Eventually, statistical time domain features were computed to capture essential information about limb movement-related signals' temporal dynamics and statistical properties. The features include mean, standard deviation, variance, root mean square, absolute difference signal, skewness, and kurtosis, as depicted in Table I. Data standardization was applied with standard scalar after constructing the tabular feature set. It centers and scales the data so that the mean is 0 and the standard deviation is 1.

TABLE I
THE LIST OF STATISTICAL AND SPECTRAL FEATURES EXTRACTED FROM EACH EEG CHANNEL.

| Feature Names | Frequency Range | Feature Names | Equations |
|---|---|---|---|
| Delta | 0.5-4 Hz | Mean | $\mu = \frac{1}{n}\sum_{i=1}^{n} x_i$ |
| Theta | 4-8 | Standard Deviation | $\sigma = \sqrt{\sum_{i=1}^{n}(x_i - \mu)^2}$ |
| Alpha) | 8-13 | Variance | $\sigma^2 = \sum_{i=1}^{n}(x_i - \mu)^2$ |
| Low Beta | 13-20 | Root Mean Square (RMS) | $RMS = \sqrt{\frac{1}{n}\sum_{i=1}^{n} x_i^2}$ |
| Mid Beta | 20-26 | Absolute Difference Signal | $ABS = |x_i - x_{i-1}|$ |
| High Beta | 26-35 | Skewness | $E\left(\frac{x-\bar{x}}{\sigma}\right)^3$ |
| -- | -- | Kurtosis | $E\left(\frac{x-\bar{x}}{\sigma}\right)^4$ |

3) *Feature Selection*

To determine the optimal number of features, we introduced a hybrid feature selection technique, as illustrated in Fig. 4. The feature selection process was divided into three steps.

In the 1$^{st}$ step, mutual information-based feature selection was applied. Mutual information (MI) *I(X; Y)* measures the mutual dependence between two variables, i.e., *X* and *Y*. Here, *X* represents the independent variable/feature matrix and *Y* represents the dependent variable or target variable. It is computed for each feature matrix with respect to the target variable.

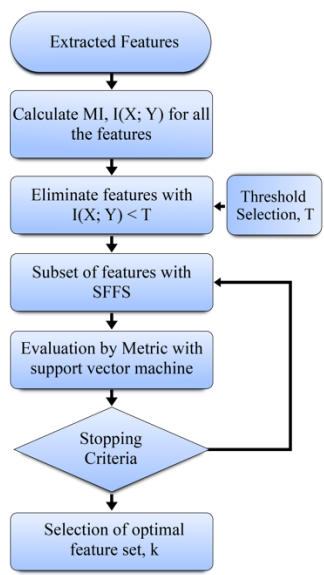

Fig. 4. An overview of our hybrid feature selection procedure. The extracted features are used to calculate individual mutual information. Then, the features are dropped based on a predefined threshold, and finally, the selected features are fed into the SFFS algorithm, which finds the optimal set of features with the help of SVM.

The mutual information between those two was calculated as follows:
$$I(X;Y) = H(X) - H(X|Y) \qquad (6)$$
Here, *H(X)* represents the entropy of the feature matrix *X*, and *H(X, Y)* is the combined/joint entropy of *X* and *Y*. The most informative features were selected by defining a threshold (*T*). Here, *T* is the minimum value of mutual information for which a feature is considered to be informative. Any value of mutual information below this is considered less informative and the features that hold the MI value below this were rejected. Mathematically, for a given feature $X_i$ and a defined threshold *T*, if $I(X_i;Y) < T$, then $X_i$ was removed. We set the threshold to *T*=0.03 and kept all the features that were greater than the threshold.

In the second step, we included SFFS-based feature selection. SFFS takes the set of all features, $Y = [y_1, y_2 ... y_d]$ of d-dimensions as input taken from the output from the 1$^{st}$ step, and it returns the final output as a subset of features $X_k = [x_j \mid j = 1, 2, ..., k; x_j \in Y]$, where the number of selected feature $k = (1, 2, ..., d)$. The size of the returned output from the SFFS algorithm is a subset of all the features and $k < d$. The algorithm initializes at $X_0 = \Phi$ and $k = 0$, which denotes an empty set, also known as a "null set." Then, we did sequential inclusion of each feature. The algorithm selected a feature $x^+$ from the remaining feature space that maximized the criterion function $J(X_{k+x})$, where *x* belongs to the set of unselected features $Y - X_k$. The selected feature was then added to the current subset $(X_k)$ and *k* was incremented by one. Then, the algorithm proceeded to step 2.

In the third step, the conditional exclusion of features was applied. The algorithm evaluates whether removing a feature $x^-$ from the current subset $(X_k)$ would lead to an improvement in performance, as determined by the criterion function$(X_k - k)$, where *x* belongs to the current subset $(X_k)$. If $J((X_k - x) > J((X_k)$, indicating a potential performance gain, the feature $x^-$ was removed from the subset $(X_k)$ and *k* is decremented by one. The algorithm then returns to step 1. If no improvement can be made or if the subset size *k* reaches a minimum value of 2 (meaning the algorithm has reduced the feature set to only two features), the algorithm terminates. It implied that no performance improvement could be achieved further. The overall approach is outlined in Algorithm 1.

ALGORITHM I
MODIFIED SEQUENTIAL FORWARD FLOATING SELECTION

**Step 1: Feature elimination with Mutual Information**
**For**
$$I_i(X_i;Y_i) = H(X_i) - H(X|Y_i)$$
**If**
$$I_i > T \text{ then } S = S + S_i$$
**end For**
Initialize $Y = \{S_1, S_2, S_3 ... S_d\}$, d=no of selected feature
While the stop criterion has not been fulfilled
**Step 2 (Inclusion)**
$$x^+ = \arg\max J(X_k + x^+), where\ x \in Y - X$$
$$K = k + 1$$
$$J_{max} = \max(J(X_{k+1}), J_{max})$$
**Step 3 (Exclusion)**
$$x^- = \arg\max J(X_k - x^-), where\ x \in X_k$$
**If**
$$J(X_k - x^-) > J(X_k):$$
Then
$$X_{k-1} = X_k - x^-$$
$$K = k - 1$$
**Go to step 3**
**Else**
**Go to step 2**

4) *Model Architecture and Hyper-parameter tuning*

We applied a multi-layer perceptron of a fully connected neural network as a classifier. Keras Tuner's [34] was used to estimate different hyper parameters, as shown in Table II. We tuned every model for every one-*vs*-one task separately. Among these, the best hyper parameters were selected based on randomly constructing neural networks with different combinations of these hyper parameters. The evaluations were done with a holdout validation of 80% training and 20% testing on all data. A random search was conducted in 100 trials for each classification task. The number of units in the hidden layers was set to be selected by random search between 20 to 30 neurons. To prevent over-fitting our model, we utilized dropout regularization. It randomly sets a fraction of the units (neurons) of a specific layer to zero during each training step. We set the dropout for random search to be chosen any in the range of 0.1 (10%) to 0.9 (90%). We applied 'ReLU' and 'LeakyReLU' activation functions for the hidden layers and 'sigmoid' for the output layer. The Adam and RMSprop-based optimizers were deployed in the hidden layer to optimize the cost function.

We incorporated binary cross entropy as a loss function of our deep learning model. The loss function measures the dissimilarity between predicted probabilities and actual binary labels (0 or 1) for each sample. Mathematically, it is defined as

$$L(y, \hat{y}) = -[y log(\hat{y}) + (1 - y) \log(1 - \hat{y})] \quad (10)$$

Where, $y$ is the actual binary label of the sample and $\hat{y}$ is the predicted probability of the samples which belong to class 1. An $L2$ regularization was employed to control the complexity of the weights. We kept the regularization strength (or lambda) of 0.01. Table II shows the estimated outputs for each hyper parameter across all tasks.

TABLE II
ESTIMATED ARCHITECTURE AND THEIR HYPER PARAMETERS FOR EACH TASK

| Task | No of hidden layers | Dropout | Units | Activation function | Optimizer |
|---|---|---|---|---|---|
| I | 2 | 0.1 | 28 | LeakyReLU | Adam |
|   |   | 0.5 | 27 | LeakyReLU |   |
| II | 1 | 0.2 | 25 | ReLU | RMSprop |
| III | 2 | 0.6 | 26 | LeakyReLU | Adam |
|   |   | 0.9 | 10 | LeakyReLU |   |
| IV | 2 | 0.2 | 26 | ReLU | Adam |
|   |   | 0.9 | 26 | LeakyReLU |   |
| V | 1 | 0.4 | 23 | ReLU | RMSprop |
| VI | 2 | 0.1 | 26 | ReLU | Adam |
|   |   | 0.6 | 8 |   |   |

*C. Performance Evaluation*

We implemented a leave one subject out cross-validation technique to obtain the performance of our proposed method. The accuracy of the model was tested with one subject's data and the rest were used for training. The accuracy was used as the performance metric to measure the model performance.

III. RESULTS

This study focused on using EEG signals to identify four distinct motor imagery tasks: left-hand movement (task I), right-hand movement (task II), feet movement (task III), and tongue movement (task IV). Further, each of the paired classification tasks is divided into six categories, as shown in Table III.

TABLE III
EACH PAIRWISE CLASSIFICATION DENOTED AS TASKS

| Task No. | Classification |
|---|---|
| Task I | Left hand vs. Right hand |
| Task II | Left hand vs. Foot |
| Task III | Left hand vs. Tongue |
| Task IV | Right hand vs. Foot |
| Task V | Right hand vs. Tongue |
| Task VI | Foot vs. Tongue |

The average accuracy of motor imagery classification using the proposed feature selection method for Task I, II, III, IV, V and VI is 73.19%, 87.48%, 65.21%, 86.86%, 83.05% and 82.38%, respectively. The classification accuracy of each subject using an optimal number of features is illustrated in Table IV. The average classification accuracy for all tasks without feature selection, with features selected by mutual information and with the proposed feature selection technique is 69.68 ± 3.21, 76.85 ± 5.64, and 79.69 ± 7.98, respectively. The average classification accuracy for each task without feature selection, using mutual information-based feature selection and using proposed feature selection, is demonstrated in Fig. 5. In contrast, Fig. 6 illustrates the average accuracies of each task for individual subjects for different sets of features.

TABLE IV
VALIDATION ACCURACIES OBTAINED FROM OPTIMAL NUMBER OF FEATURES.

| SUB | Task I | Task II | Task III | Task IV | Task V | Task VI |
|---|---|---|---|---|---|---|
| 1 | 65.71 | 82.14 | 65.95 | 83.8 | 65.73 | 63.63 |
| 2 | 65.03 | 78.32 | 65.73 | 80.55 | 72.22 | 72.22 |
| 3 | 72.72 | 84.5 | 55.24 | 79.72 | 85.41 | 76.92 |
| 4 | 78.41 | 89.99 | 65.46 | 81.11 | 78.87 | 73.42 |
| 5 | 75.17 | 80.98 | 70.62 | 89.36 | 89.43 | 88.11 |
| 6 | 64.58 | 86.8 | 65.73 | 85.41 | 81.11 | 87.41 |
| 7 | 84.5 | 96.47 | 67.13 | 98.57 | 94.32 | 97.16 |
| 8 | 84.5 | 95.77 | 66.43 | 97.85 | 92.9 | 97.16 |
| 9 | 68.05 | 92.36 | 64.58 | 85.41 | 87.5 | 85.41 |
| AVG | 73.19 | 87.48 | 65.21 | 86.86 | 83.05 | 82.38 |
| STD | 7.98 | 6.55 | 4.11 | 7.08 | 9.54 | 11.56 |

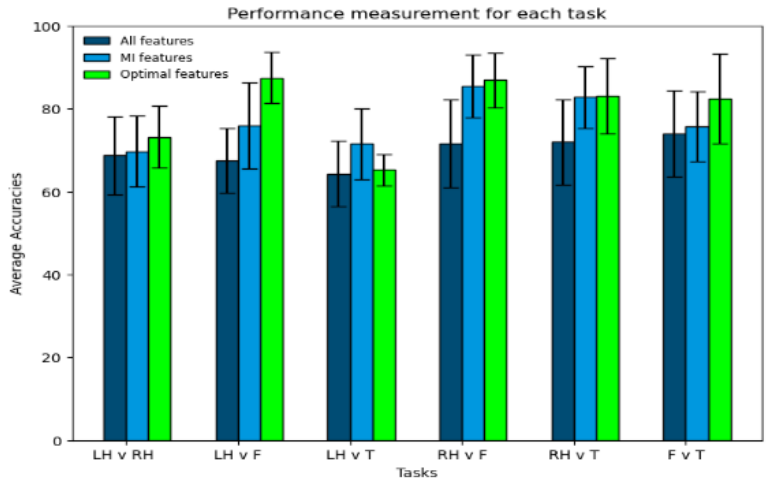

Fig. 5. The average accuracies of each classification task across all the subjects are depicted with the help of a comparative bar plot comparing three sets of features. The error bar is also included to show the deviation of accuracies for each.

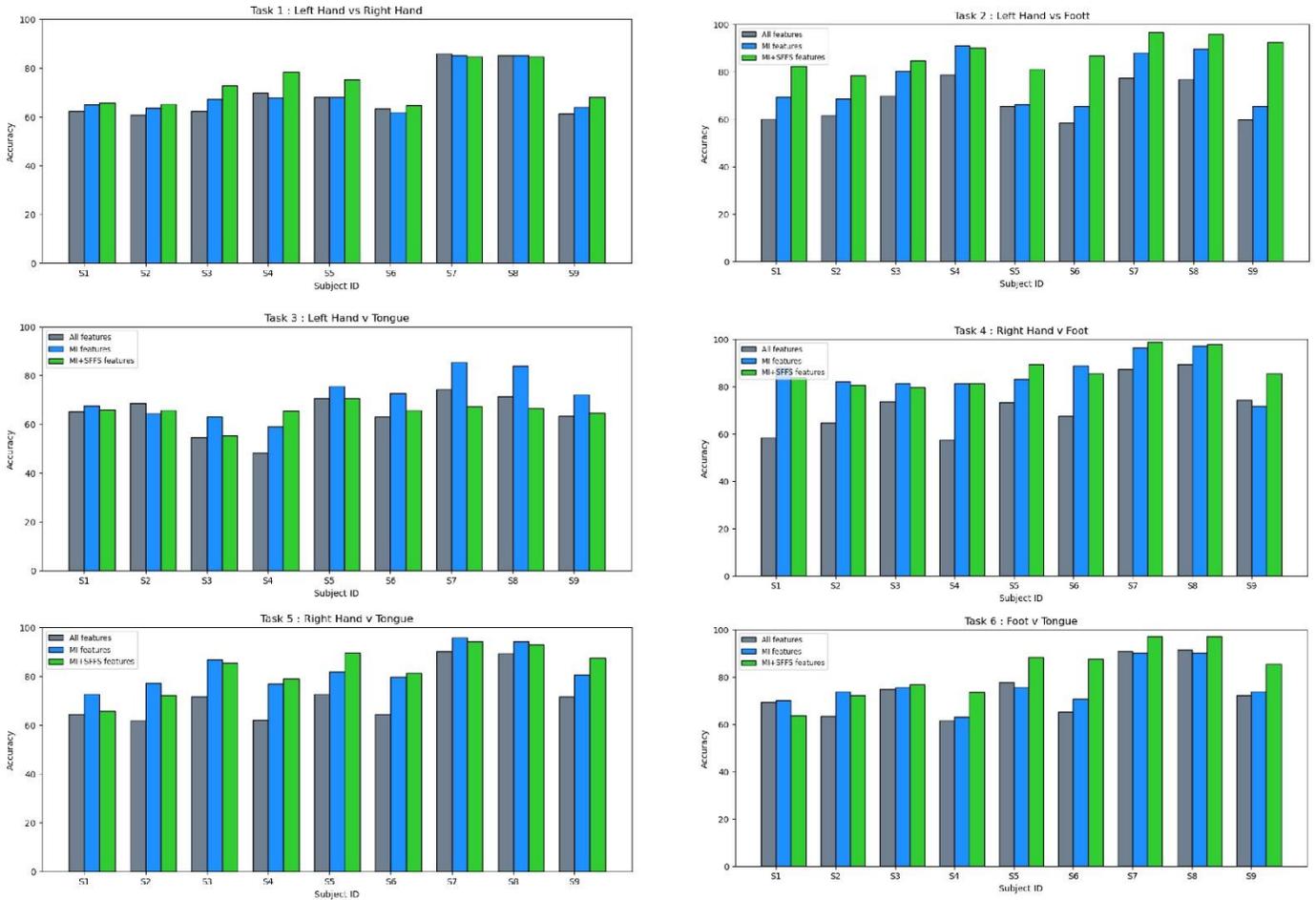

Fig. 6. The obtained best-performing accuracy for each task was compared among all the features compared with features which are selected manually with higher mutual gain and the optimal features with respect to each the subjects. The proposed feature selection techniques outperformed all tasks except task 3 (LH v T).

The classification performance incorporated with our proposed hybrid feature selection techniques was compared with the state-of-the-art- feature selection techniques-based work in Table V. Our proposed approach outperformed the existing alternatives for all tasks except Task III.

TABLE V
PERFORMANCE COMPARISON OF THE DIFFERENT FEATURE SELECTION ALGORITHMS IN CLASSIFYING DIFFERENT MOTOR IMAGERY TASKS.

| FEATURE SELECTION | TASK I | TASK II | TASK III | TASK IV | TASK V | TASK VI | AVG |
|---|---|---|---|---|---|---|---|
| MUTUAL INFORMATION | 69.73 | 75.89 | **71.48** | 85.46 | 82.82 | 75.76 | 76.85 |
| MRMR | 71.27 | 62.26 | 65.87 | 64.47 | 64.66 | 66.34 | 65.81 |
| RFE-SVM | 70.71 | 63.82 | 70.02 | 62.49 | 66.05 | 70.03 | 67.18 |
| FIREFLY ALGORITHM | 71.87 | 67.79 | 62.49 | 65.89 | 70.76 | 72.42 | 68.53 |
| SFS-SVM | 70.63 | 81.88 | 65.73 | 62.79 | 65.16 | 75.09 | 70.21 |
| **PROPOSED** | **73.19** | **87.48** | 65.21 | **86.86** | **83.05** | **82.38** | **79.70** |

In the initial stage of feature selection with mutual information, the choice of suitable threshold is a challenging task. We empirically selected the threshold, which was 0.03. The performance of the model varies with the variation of the threshold, which is shown in Fig. 7.

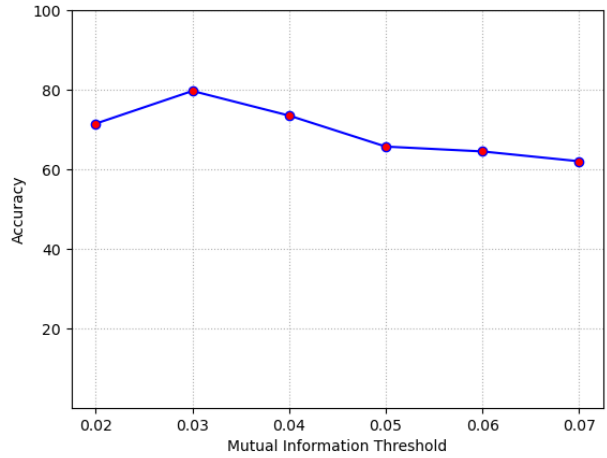

Fig. 7. Average accuracy obtained using different thresholds of mutual information.

The proposed hybrid feature selection method also reduces the high dimensionality of features while working with large number of feature set. From the total set of extracted features, the number of features selected in step 1 and step 2 is compared in Fig. 8. According to the figure, the first step achieves a significant reduction in features, around half of total number of features.

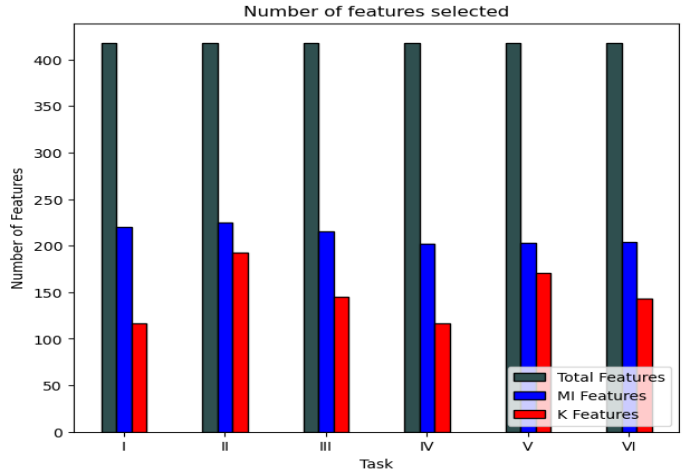

Fig. 8. A comparison for the number of features that were selected by 1st stage i.e., mutual information vs. 2nd stage which was SFFS for all the six tasks.

## IV. Discussion

We propose a hybrid approach that combines two distinct feature selection techniques, leverages the robustness of those feature selection techniques, and enhances classification accuracy. In the initial stage, feature selection based on mutual information is employed to decrease additional dimensions and eliminate redundant features. Subsequently, the SFFS method identifies the optimal subset of features from those selected in the first stage. Since redundant features have already been eliminated in the initial stage, the SFFS algorithm sequentially explores various combinations of features without needing to select from a broader range. Thus, reducing the dimensionality. The proposed approach of feature selection is reliable in selecting the best subset of features from a diverse set of features. EEG recordings capture brain activity through multiple channels. Each channel records a unique signal, differing in amplitude, frequency, signal-to-noise ratio, and other characteristics. There's a chance that a single set of features from a few channels cannot capture all effective information for motor imagery classification. Employing a large number of features relevant to EEG signals can solve this problem but increases the computational cost due to the high dimensionality of the feature sets. Since our proposed approach uses multiple steps to reduce dimensionality to find an optimal number of features from a large pool of features, it is well suited to handle large feature sets.

As shown in Fig. 9, when locating the optimal set, the performance of task I and task III does not differ significantly for different sets of features, and several sets of features perform almost as well as the best optimal set. With the help of proposed feature selection scheme, the most contributing channels was estimated by comparing the number of features that were selected from each channel to the maximum number of features that were selected by a channel for that specific classification task. The following mathematical expression was used to visualize the saliency maps.

$$Channel\ Significance = k_i / k_m \qquad (6)$$

Where, $k_i$ is the number of features selected for the $i^{th}$ channel, and $k_m$ is the maximum number of features that are selected for the $m^{th}$ channel. We have plotted these estimated values of the spatial significance of each channel with a heat map and the location of channels in the scalp. All the saliency maps are shown in Fig. 10, which indicates the significance of channels. For example, Fig. 9b shows the highest number of features selected from the right side of the scalp and from the middle and front. Therefore, the information associated with the channel connected to its respective region exerts the most significant impact on classifying samples between the left hand and foot. Analysis of Table IV reveals that Tasks II and IV achieved higher accuracy compared to other classification tasks. This suggests a potential correlation between higher accuracy and the ability to identify optimal channel priorities.

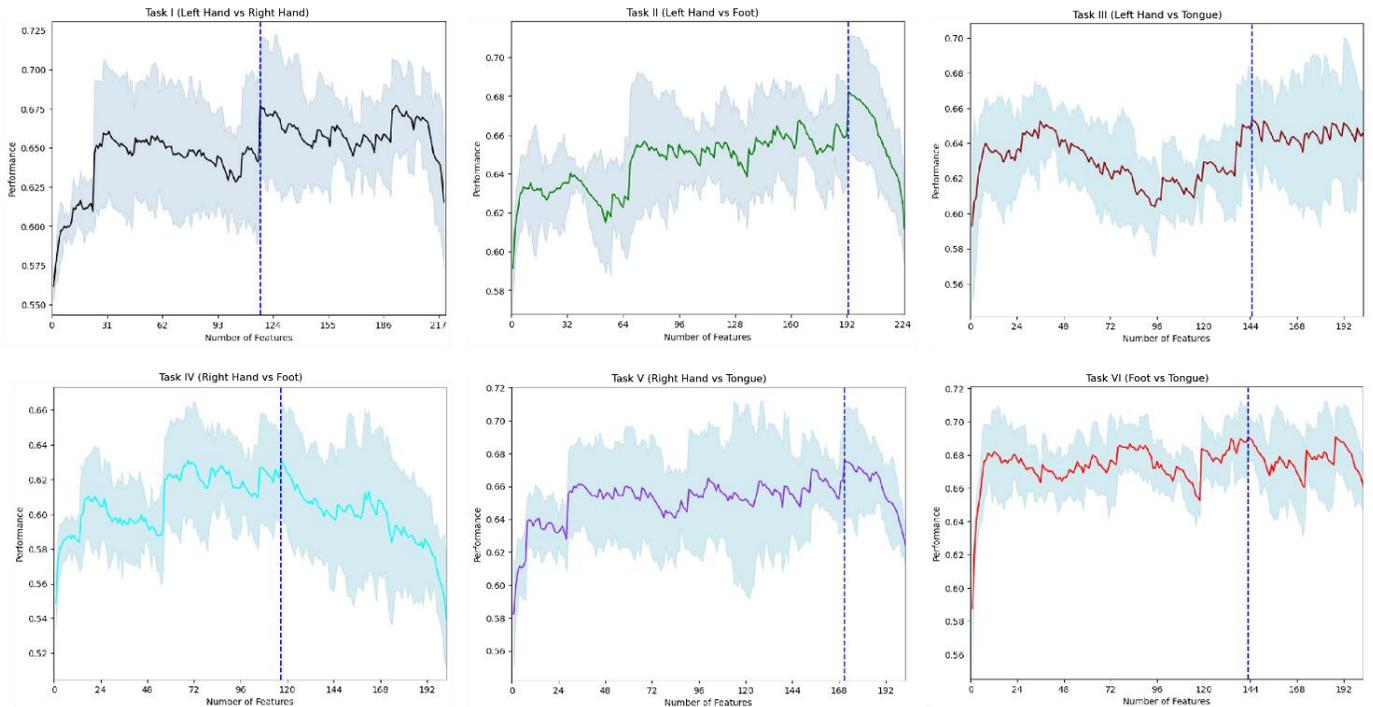

Fig. 9. The optimal feature set is the final set of features that were selected for each classification task. The number of features that are chosen with respect to each channel implies the contribution of that channel to that specific channel.

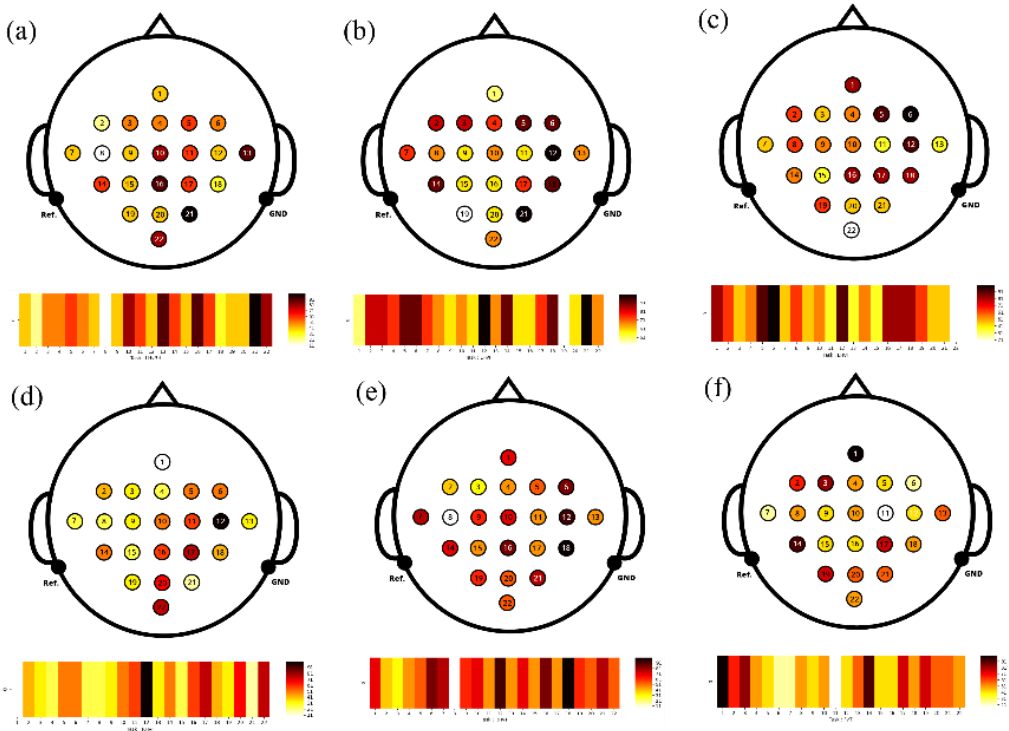

Fig. 10. The optimal feature set is the final set of features that were selected for each classification task. The number of features that are selected with respect to each channel implies the contribution of that channel for that specific classification task. Here, for one task, the frequency of some channels is higher than that of other channels which indicates their importance for that task. Here, the number of selected features of each channel is shown on the heat map and also on the headset. The heat map depicts the priority of each value, with darker colors indicating higher priority, which means a greater number of features are selected from that channel. The six-heat map and illustrated salient maps depict a) Task I, b) Task II, c) Task III, d) Task IV, e) Task V, and f) Task VI.

The comparison of the proposed work with existing literature for motor imagery classification is demonstrated in Table VI. As shown in Table VI, our model outperforms most of the reported results where BCIC-IV 2a benchmark dataset was used. The accuracy reported in [28] [29] [30] were slightly higher than our model. This might be due to the use of different datasets that led to this kind of performance variation.

TABLE VI
THE COMPARISON OF THE PROPOSED WORK WITH THE EXISTING WORK IN THE LITERATURE FOR MOTOR IMAGERY CLASSIFICATION USING EEG signal

| Study | Dataset | Feature | Classifier | Acc |
| --- | --- | --- | --- | --- |
| Lawhern et al. [18] | BCIC IV 2a | Time-series | CNN | 67.00 |
| Luo et al. [19] | BCIC IV 2a | Spatial-frequency | RNN-GRU | 73.56 |
| Wang et al. [16] | BCIC IV 2a | Calculated features | LSTM | 70.77 |
| Hassanpour et al. [21] | BCIC IV 2a | Time-series | SAE | 71.08 |
| Yao Guo et. al. [30] | BCIC III 4a | FCCSP | LDA | 82.01 |
| A. Liu *et al.* [27] | BCIC IV 2a | Calculated Features | SRDA | 70.20 |
| Jing Jiang et al. [29] | BCIC (III 3a & 4a, IV-1) | CSP+LASSO | SVM | 88.58 |
| X. Yu et al. [24] | BCIC III 4a | CAR, LAP, CSP + PCA | SVM | 74.59 |
| N.S. Malan & S. Sharma [28] | BCIC (II-3 & IV-2b) | DTCWT+RNCA | SVM | 80.70 |
| **Proposed Work** | BCIC IV 2a | Optimized Features | FCNN | **79.70** |

## V. Conclusion

This work presents a hybrid feature selection technique to select features from a diverse set of EEG features and utilize those selected features with a shallow neural network architecture for motor imagery classification. The proposed method can effectively discriminate between two different motor planning scenarios with an average accuracy of 79.70% and a standard deviation of 7.98%. The performance of our proposed model with different feature selection scenarios outperforms the existing state-of-the-art feature selection alternatives for EEG feature selection for motor imagery classification. These findings can potentially advance the development of different motor imagery applications using EEG signals.

15